\documentclass[12pt]{article}%
\usepackage{amsmath}
\usepackage{amsfonts}
\usepackage{amssymb}
\usepackage{graphicx}%
\begin{document}
\begin{titlepage}
\begin{flushright}
PUPT-2191\\
hep-th/0602011
\end{flushright}
\vspace{7 mm}
\begin{center}
\huge{Beyond Space-Time}
\end{center}
\vspace{10 mm}
\begin{center}
{\large
A.M.~Polyakov\\
}
\vspace{3mm}
Joseph Henry Laboratories\\
Princeton University\\
Princeton, New Jersey 08544
\end{center}
\vspace{7mm}
\begin{center}
{\large Abstract}
\end{center}
\noindent
These notes, based on the remarks made at the 23 Solvay Conference, collect several
speculative ideas concerning gauge/ strings duality, de Sitter spaces,dimensionality and the cosmological
constant
\vspace{7mm}
\begin{flushleft}
February 2006
\end{flushleft}
\end{titlepage}

Let me begin with some general views. In fundamental physics we invent
dynamical mechanisms, based on the first principles, like the Higgs mechanism,
and construct models, based on these mechanisms. In the case of the standard
model this path led to a tremendous success. Finding the right mechanisms may
be easier than constructing a detailed model. For example, the Yang-Mills
theory was at first misapplied to describe the rho mesons, and only ten years
later found its right place in QCD. This is a kind of danger we should be
aware of.

Another danger is to get distracted by non-dynamical anthropic arguments,
which recently acquired some popularity. I find the anthropic principle
irrelevant. It is unlikely to uncover fundamental ideas and equations
governing the universe. But, in spite of these misanthropic remarks, I believe
that in special cases anthropic arguments may be appropriate.

In what follows I shall briefly describe various mechanisms operating in and
around string theory. This theory provides a novel view of space-time. I would
compare it with the view of heat provided by statistical mechanics. At the
first stage the word "heat" describes our feelings. At the second we try to
quantify it by using equations of thermodynamics. And finally comes an
astonishing hypothesis that heat is a reflection of molecular disorder. This
is encoded in one of the most fascinating relations ever, the Boltzmann
relation between entropy and probability.

Similar stages can be discerned in string theory. The first is of course the
perception of space-time. The second is its description using the Einstein
equations. The third is perhaps a possibility to describe quantum space-time
by the boundary gauge theory. Let us discuss in more details our limited but
important knowledge of the gauge/string correspondence.

\section{ Gauge /String correspondence}

It consists of several steps. First we try to describe the dynamics of a
non-abelian flux line by some string theory. That means, among other things
that the Wilson loop $W(C)$ must be represented as a sum over 2d random
surfaces immersed in the flat 4d space-time and bounded by the contour $C.$
Surprisingly, strings in 4d behave as if they are living in the 5d space, the
fifth (Liouville) dimension being a result of quantum fluctuations\cite{p81} .
More detailed analyses shows that while the 4d space is flat, the 5d must be
warped with the metric%
\begin{equation}
ds^{2}=d\varphi^{2}+a^{2}(\varphi)d\overrightarrow{x}^{2}%
\end{equation}
where the scale factor $a(\varphi)$ must be determined from the condition of
conformal symmetry on the world sheet \cite{p97}. This is the right habitat
for the gauge theory strings. If the gauge theory is conformally invariant
(having a zero beta function) the isometries of the metric must form a
conformal group. This happens for the space of constant negative curvature,
$a(\varphi)\sim\exp c\varphi$ \cite{m97}. The precise meaning of the
gauge/strings correspondence \cite{gkp98},\cite{w98} is that there is an
isomorphism between the single trace operators of a gauge theory, e.g.
$Tr(\nabla^{k}F_{\mu\nu}\nabla^{l}F_{\lambda\rho}...)$ and the on-shell vertex
operators of the string, propagating in the above background. In other words,
the $S-$ matrix of a string in the 5d warped space is equal to a correlator of
a gauge theory in the flat 4d space. The Yang -Mills equations of motion imply
that the single trace operators containing $\nabla_{\mu}F_{\mu\nu}$ are equal
to zero. On the string theory side it corresponds to the null vectors of the
Virasoro algebra, leading to the linear relations between the vertex
operators. If we pass to the generating functional of the various Yang- Mills
operators, we can encode the above relation in the formula%
\begin{equation}
\Psi_{WOE}[h_{\mu\nu}(x),...]=\langle\exp\int dxh_{\mu\nu}(x)T_{\mu\nu
}(x)+...\rangle_{Y.-M}%
\end{equation}
Here at the left hand side we have the "wave function of everything" (WOE). It
is obtained as a functional integral over 5d geometries with the metric
$g_{mn}(x,y)$, where $y=\exp-c\varphi,$ satisfying asymptotic condition at
infinity $(y\rightarrow0)$ $g_{\mu\nu}\rightarrow\frac{1}{y^{2}}(\delta
_{\mu\nu}+h_{\mu\nu}(x)).$ It differs from the "wave function of the universe
" by Hartle and Hawking only by the $y^{-2}$ factor. On the right side we have
an expression defined in terms of the Yang- Mills only, $T_{\mu\nu}$ being its
energy- momentum tensor. The dots stand for the various string fields which
are not shown explicitly. An interesting unsolved problem is to find the wave
equation satisfied by $\Psi.$ It is not the Wheeler -de Witt equation. The
experience with the loop equations of QCD tells us that the general structure
of the wave equation must be as following%
\begin{equation}
\mathcal{H}\Psi=\Psi\ast\Psi
\end{equation}
where $\mathcal{H}$ is some analogue of the loop Laplacian and the star
product is yet to be defined. This conjectured non-linearity may lead to the
existence of soliton-like WOE-s.

The formula (2 ) , like the Boltzmann formula, is relating objects of very
different nature. This formula has been confirmed in various limiting cases in
which either LHS or RHS or both can be calculated. I suspect that, like with
the Boltzmann formula, its true meaning will still be discussed a hundred
years from now.

\section{ de Sitter Space and Dyson's instability}

Above we discussed the gauge/ strings duality for the geometries which
asymptotically have negative curvature. What happens in the de Sitter case ?
It is not very clear. There have been a number of attempts to understand it
\cite{str01}. We will try here a different approach. It doesn't solve the
problem, but perhaps gives a sense of the right direction.

Let us begin with the 2d model, the Liouville theory. Its partition function
is given by%
\begin{equation}
Z(\mu)=\int D\varphi\exp\{-\frac{c}{48\pi}\int d^{2}x(\frac{1}{2}%
(\partial\varphi)^{2}+\mu e^{\varphi})
\end{equation}
For large $c$ (the Liouville central charge) one can use the classical
approximation. The classical solution with positive $\mu$ describes the AdS
space with the scalar curvature -$\mu.$By the use of various methods
\cite{kpz} one can find an exact answer for the partition function, $Z\sim
\mu^{\alpha}$ where $\alpha=\frac{1}{12}[c-1+\sqrt{(c-1)(c-25)}].$ In order to
go to the de Sitter space we have to change $\mu\Rightarrow-\mu.$ Then the
partition function acquires an imaginary part, $\operatorname{Im}Z\sim\sin
\pi\alpha|\mu|^{\alpha}$ . It seems natural to assume that the imaginary part
of the Euclidean partition function means that the de Sitter space is
intrinsically unstable. This instability perhaps means that due to the Gibbons
-Hawking temperature of this space it "evaporates" like a simple black hole.
In the latter its mass decreases with time, in the de-Sitter space it is the
cosmological constant. If we define the Gibbons- Hawking entropy $S$ in the
usual way, $S=(1-\beta\frac{\partial}{\partial\beta})\log Z,$ we find another
tantalizing relation, $\operatorname{Im}Z\sim e^{S},$which holds in the
classical limit, $c\rightarrow\infty.$ Its natural interpretation is that the
decay rate of the dS space is proportional to the number of states, but it is
still a speculation, since the precise meaning of the entropy is not clear.
For further progress the euclidean field theory, used above, is inadequate and
must be replaced with the Schwinger- Keldysh methods.

In higher dimensions we can try once again the method of analytic continuation
from the AdS space. The AdS geometry is dual to a conformally invariant gauge
field theory. In the strong coupling limit (which we consider for simplicity
only ) the scalar curvature of the AdS, $R\propto\frac{1}{\sqrt{\lambda}}$ (
$\lambda=g_{YM}^{2}N$ ). So, the analytic continuation we should be looking
for is $\sqrt{\lambda}\Rightarrow-\sqrt{\lambda}$. In order to understand what
it means in the gauge theory, let us notice that in the same limit the Coulomb
interaction of two charges is proportional to $\sqrt{\lambda}$ \cite{m98}.
Hence under the analytic continuation we get a theory in which the same
charges attract each other. Fifty years ago Dyson has shown that the vacuum in
such a system is unstable due to creation of the clouds of particles with the
same charge. It is natural to conjecture that Dyson's instability of the gauge
theory translates into the intrinsic instability of the de Sitter space. Once
again the cosmological constant evaporates.

\section{Descent to four dimensions}

Critical dimension in string theory is ten . How it becomes four ? If we
consider type two superstrings, the 10d vacuum is stable, at least
perturbatively., and stays 10d. Let us take a look at the type zero strings,
which correspond to a non-chiral GSO projections. These strings contain a
tachyon, described by a relevant operator of the string sigma model. Relevant
operators drive a system from one fixed point to another. According to
Zamolodchikov's theorem, the central charge must decrease in the process. That
means that the string becomes non-critical and the Liouville field must
appear. The Liouville dimension provides us with the emergent "time" in which
the system evolves and changes its effective dimensionality (the central
charge). As the "time" goes by, the effective dimensionality of the system
goes down. If nothing stops it, we should end up with the $c=0$ system which
has only the Liouville field. It is possible,however, that non-perturbative
effects would stop this slide to nothingness \cite{p86}. In four dimensions we
have the $B$ -field instantons, described by the formula (at large distances)
($dB)_{\mu\nu\lambda}=q\epsilon_{\mu\nu\lambda\rho}\frac{x_{\rho}}{x^{4}}$ .
In the modern language they correspond to the NS5 branes. These instantons
form a Coulomb plasma with the action $S\sim\sum\frac{q_{i}q_{j}}{(x_{i}%
-x_{j})^{2}}.$ As was explained in \cite{p86} , the Debye screening in this
plasma causes "string confinement", turning the string into a membrane.
Formally this is described by the relation%
\begin{equation}
\langle\exp i\int B_{\mu\nu}d\sigma_{\mu\nu}\rangle\sim e^{-aV}%
\end{equation}
where we integrate over the string world sheet and $V$ is the volume enclosed
by it. There is an obvious analogy with Wilson's confinement criterion. While
the gravitons remain unaffected, the sigma model description stops being
applicable and hopefully the sliding stops at 4d.

\section{Screening of the cosmological constant}

Classical limits in quantum field theories are often not straightforward. For
example, classical solutions of the Yang - Mills theory describing interaction
of two charges have little to do with the actual interaction. The reason is
that because of the strong infrared effects the effective action of the theory
has no resemblance to the classical action. In the Einstein gravity without a
cosmological constant the IR effects are absent and the classical equations
make sense. This is because the interaction of gravitons contain derivatives
and is irrelevant in the infrared.

The situation with the cosmological term is quite different, since it doesn't
contain derivatives. Here we can expect strong infrared effects \cite{p82} ,
see also \cite{jac05} for the recent discussion.

Let us begin with the 2d model ( 3) . The value of $\mu$ in this lagrangian is
subject to renormalization. Perturbation theory generates logarithmic
corrections to this quantity. It is easy to sum up all these logs and get the
result $\mu_{ph}=\mu(\frac{\Lambda}{\mu_{ph}})^{\beta}$ , with $\beta=\frac
{1}{12}[c-13-\sqrt{(c-1)(c-25)}].$Here $\Lambda$ is an UV cut-off while the
physical (negative) cosmological constant $\mu_{ph}$ provides a
self-consistent IR cut-off. We see that in this case the negative cosmological
constant is anti-screened.

In four dimensions the problem is unsolved. For a crude model one can look at
the IR effect of the conformally flat metrics. If the metric $g_{\mu\nu
}=\varphi^{2}\delta_{\mu\nu}$ is substituted in the Einstein action $S$ with
the cosmological constant $\Lambda,$the result is $S=\int d^{4}x$[-$\frac
{1}{2}(\partial\varphi)^{2}+\Lambda\varphi^{4}].$ There is the well known
non-positivity of this action. This is an interesting topic by itself, but
here we will not discuss it and simply follow the prescription of Gibbons and
Hawking and change $\varphi\Rightarrow i\varphi.$ After that we obtain a well
defined $\varphi^{4}$ theory with the coupling constant equal to $\Lambda.$
This theory has an infrared fixed point at zero coupling, meaning that the
cosmological constant screens to zero.

There exists a well known argument against the importance of the infrared
effects. It states that in the limit of very large wave length the
perturbations can be viewed as a change of the coordinate system and thus are
simply gauge artefacts.This argument is perfectly reasonable when we discuss
small fluctuations at the fixed background (see \cite{wood} for a different
point of view). However in the case above the effect is non-perturbative- it
is caused by the fluctuation of the metric near zero, not near some
background. In this circumstances the argument fails. Indeed, if we look at
the scalar curvature, it has the form $R\sim\varphi^{-3}\partial^{2}\varphi.$
We see that while for the perturbative fluctuations it is always small because
of the second derivatives, when $\varphi$ is allowed to be near zero this
smallness can be compensated. In the above primitive model the physical
cosmological constant is determined from the equation $\Lambda_{ph}%
=\frac{const}{\log\frac{1}{\Lambda_{ph}}}$ which always has a zero solution.
One would expect that in the time-dependent formalism we would get a slow
evaporation instead of this zero. The main challenge for these ideas is to go
beyond the conformally flat fluctuations. Perhaps gauge/ strings
correspondence will help.

As it is clear from the list of the references below, these ideas (except for
the gauge/strings correspondence) did not attract any attention. Perhaps they
don't deserve it. My best hope, however, is that some of them may serve as
small building blocks of the future theory, the vague contours of which we can
discern at the horizon.

I am deeply grateful to Thibault Damour for many years of discussions on the
topics of this paper. I would also like to thank David Gross and Igor Klebanov
for useful comments.

This work was partially supported by the NSF grant 0243680. Any opinions,
findings and conclusions or recommendations expressed in this material are
those of the authors and do not necessarly reflect the views of the National
Science Foundation.

\end{document}